\documentclass[aps,prl,twocolumn,superscriptaddress,showpacs]{revtex4}

\usepackage{graphicx}
\DeclareGraphicsExtensions{.eps}
\begin{document}
\title{Infrared emission spectrum and potentials of $0_{u}^{+}$ and $0_{g}^{+}$ states of Xe$_{2}$ excimers produced by electron impact}
\author{A. F. Borghesani}\email{borghesani@padova.infm.it}
\affiliation{Department of Physics, University of Padua} \affiliation{I.N.F.N., Sezione di Padova, via F. Marzolo 8, I--35131 Padua, Italy }
\author{G. Carugno}
\affiliation{I.N.F.N., Sezione di Padova, via F. Marzolo 8, I--35131 Padua, Italy }
\author{I. Mogentale}\affiliation{Department of Physics, University of Padua} \affiliation{I.N.F.N., Sezione di Padova, via F. Marzolo 8, I--35131 Padua, Italy }

\begin{abstract}
We present an 
investigation of the Xe$_{2}$ excimer emission spectrum 
observed 
in the near infrared range about 7800 cm$^{-1}$ in pure Xe gas and in an Ar (90\%) --Xe (10\%) mixture 
 and obtained by exciting the gas with energetic electrons. The Franck--Condon simulation of the spectrum shape suggests that emission stems from a bound--free molecular transition never studied before. The states involved are assigned as the bound $(3)0_{u}^{+} $ state with $6p\, [1/2]_{0}$ atomic limit and the dissociative $(1)0_{g}^{+}$ state with $6s\, [3/2]_{1}$ limit. Comparison with the spectrum simulated by using theoretical potentials shows that the dissociative one does not reproduce correctly the spectrum features. 
\end{abstract}
\maketitle
Luminescence of rare gas excimers is 
studied for applications in the vacuum ultraviolet (VUV) 
\cite{ledru2006}. It is worth recalling, for instance, the Xe--based high--energy particle detectors, in which the VUV scintillation light produced by the transit of an ionizing particle is detected \cite{knoll}. Other applications deal with the production of coherent and incoherent VUV light sources \cite{rhodes1979}. 

Xe is of great interest because of its importance as an intense source of VUV radiation. Nearly all of the investigations on Xe excimers 
aim at explaining 
the processes leading to the emission of the VUV 1st-- and 2nd continuum observed under different experimental conditions in gas excited by discharges \cite{colli1954}, UV photons \cite{
broadmann1976
,dutuit1978}, multiphotons \cite{gornik1982,dehmer1986}, or high--energy particles
\cite{koe1974,keto1974,leichner1976,
wenck1979}.

The 1st continuum at $152$ nm is due to radiative transitions from the vibrationally excited $(0_{u}^{+})_{v^{\prime}\gg 0}$ state, correlated with the resonant atomic state $5p^{5}6s\,(^{3}\!P_{1}),$ to the dissociative $0_{g}^{+}$ ground state. 
The 2nd continuum at $
170$ nm consists of the overlapping bound--free emission from the lowest 
vibrationally relaxed $(0_{u}^{-},\, 1_{u})$ 
states correlated with the metastable  
state $5p^{5}6s\,(^{3}\!P_{2})$ \cite{wenck1979,gornik1982}.

The two continua are thus produced by two different VUV emission processes whose kinetics accounts for the observation that at low pressure ($P< 2\cdot 10^{4} $ Pa) lumiscence is due to the 1st continuum, whereas it consists of the 2nd one for $P>5\cdot 10^{4}$ Pa \cite{museur1994}.

The excimer structure has been investigated theoretically with \emph{ab initio}  \cite{ermler1978,jonin2002_II} or model \cite{mulliken1970} 
calculations of the molecular potentials and experimentally by analyzing spectroscopic data \cite{keto1974,
castex1981,gornik1982,raymond1984,
koe1995}.

Metastable atomic--  and $g$ molecular states are mainly studied because of their importance for VUV emission \cite{moutard1988}. The kinetics of the processes leading to excimer formation and 
VUV emission 
has been clarified in 
spectral and time resolved experiments, in which lifetimes and rate constants are determined \cite{raymond1984,
museur1994,leichner1976,millet1978,broadmann1977,wenck1979,bonifield1980,salamero1984,
moutard1986,
gornik1982,alekseev1999,ledru2006}. 

Actually, the possibility has been neglected that, in this cascade of processes, molecular transitions occur in the infrared (IR) range. Only an IR spectrum 
centered about 800 nm was observed and attributed to a $0_{g}^{+} (6p[1/2]_{0})\rightarrow 0_{u}^{+} (6s[3/2]_{1})_{v^{\prime}\gg0}$ transition. $v^{\prime}\gg 0$ means that the final state is in a highly excited vibrational state  \cite{dehmer1986}. 

No further measurements of IR emission can be found.
Their number might be so scanty because the potential minimum of higher--lying bound excimer states occur at an internuclear distance, at which the weakly bound ground state potential is strongly repulsive, and are not easily 
reached by multiphoton selective excitation. By contrast, broad--band excitation using high--energy charged particles \cite{leichner1976} 
produces excited atoms with such high kinetic energy that can collide at short distance with ground state atoms yielding higher excimer states, although there is no control on their parity.

Recently, we 
observed for the first time a broad emission spectrum centered at $\lambda \approx \, 1.3\, \mu$m ($\tilde \nu \approx 7860 $ cm$^{-1}$) in both pure Xe gas and in a Xe(10\%)--Ar(90\%) 
mixture at room temperature by exciting the gas with a pulsed beam of 70--keV electrons   \cite{Borghesani2001}. 
Details of the tecnique can be found in literature.  
It is only worth recalling here 
 that we use a 
FT--IR spectrometer in stepscan mode. 
 An InGaAs photodiode with flat responsivity in the range $(0.6\leq\tilde\nu\leq1.2)\cdot 10^{4}$ cm$^{-1}$ is used as detector.

The FWHM of this band is $\Gamma\approx 900$ cm$^{-1}$ at $P\approx 2\cdot 10^{4}$ Pa. Its value relative to the central wave number $\tilde \nu_{m}, $ $\Gamma /\tilde\nu_{m}\approx 0.115 , $ is comparable with the value $0.116$ for the 2nd continuum  \cite{koe1974,jonin1998}.
In the limit of low $P,$ $\tilde\nu_{m}$ and $\Gamma $ are the same both in the pure gas and in the mixture. Without any further inquiries, as we were interested on the excimer interaction with the high density environment \cite{Borghesani2005},  we 
attributed the 
emission to a Xe$_{2}$ bound--free 
transition between a 
 state dissociating into the $5p^{5}6p$ manifold and one of $5p^{5}6s$ configuration \cite{Borghesani2001}.

Upon improving our experimental technique, especially using a LN$_{2}$-cooled InSb photodiode detector, we have been able to make high--resolution time--integrated measurements of the excimer IR emission that allow for the first time a more precise assignment of the molecular states involved in the transition. This goal is accomplished by comparing the observed spectrum 
with the spectrum 
calculated by means of recently published potentials 
of higher 
molecular states \cite{jonin2002_II}.
 \begin{figure}[htbp]
 \begin{center}
 \includegraphics[width=\columnwidth]{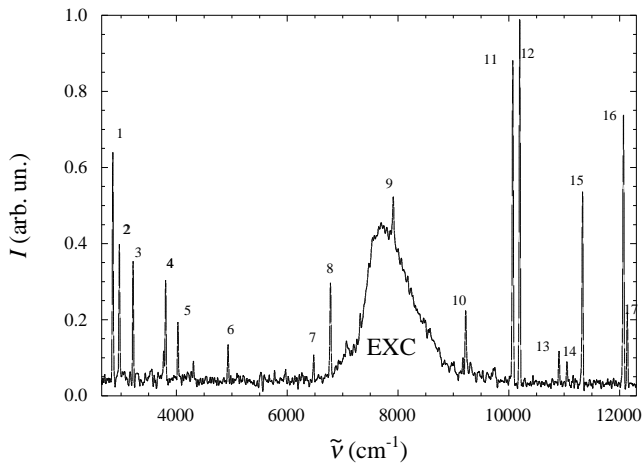}
\caption{IR emission spectrum of electron--impact excited Xe gas at $P=0.1$ MPa.The broad continuum (EXC) is the excimer spectrum. Atomic lines are numbered (see text). \label{fig:broadband}}\end{center} 
 \end{figure}

In Fig. \ref{fig:broadband} we show the IR spectrum recorded in an extended wave number range with $16$ cm$^{-1}$ resolution in pure Xe gas at $P = 0.1$ MPa and $T\approx 300$ K.
The excimer band 
appearing in the 
center 
 is surrounded by several atomic lines. An analysis of the nearby atomic transitions 
 may shed light on what atomic states the excimer is correlated with.
 Lines 1--6 stem from $5d-6p$ transitions.
Lines 7--9 are $7s-6p$ transitions. Lines 10--12 are $6p-6s$ transitions. Line 13 is an unresolved doublet $6p-6s$ and $6d-6p.$ Line 14 is a $6s-5p$ transition and lines 15--17 are $6p-5p$ transitions \cite{NIST}.

The presence of the $7s-6p$ lines very close to the excimer band suggests that the upper bound molecular state is related to the atomic $6p$ manifold. In particular, line 9 is 
due to a transition from a 
$7s\, [1/2]^{o}$ atomic state to a 
$ 6p\, [1/2]_{0}$ one \cite{NIST}. We thus assume that the 
 bound molecular state is correlated with the latter limit. 

At the density of the present experiment, $N\approx 2.4\cdot 10^{25}$ m$^{-3},$ the mean free time between collisions 
is estimated to be $\tau_{c}\approx 10^{-11}$~s \cite{Borghesani2001}, whereas the predissociation time of the bound molecular state induced by an
 avoided crossing with the $g$ molecular potential 
 related to the $
 5d\, [1/2]_{1}$ limit is estimated to be $\tau_{p}\approx 10^{-10}$ s \cite{museur1994}. The large collision rate thus leads to a quick electronic relaxation of the excimer which would otherwise predissociate. 

At present, there are no estimates for the radiative lifetime and 
 decay rate for vibrational relaxation for this bound 
 state. We assume that they 
 do not differ too much from those 
 of the states responsible of the VUV continua.
 The radiative lifetime of highly excited vibrational  states of the $0_{u}^{+}$ and $(1_{u}, \,0_{u}^{-})$ excimers is estimated to be $\tau_{1}\approx 5 $ ns \cite{moutard1988} and $\tau_{2} \approx 40$ ns \cite{keto1974}, respectively. The decay rate for vibrational relaxation $k_{3}$ is the same for all states \cite{bonifield1980} yielding a decay time $\tau_{3}=(k_{3}N)^{-1}\approx  0.65 $ ns. In any case, as $\tau_{c}$ is much shorter than all characteristic times, we assume that collisions stabilize excimers electronically and quickly establish thermal equilibrium.

The simulation of the line shape by means of Franck--Condon calculations requires the
knowledge of the potential energy curves of the initial and final molecular states and of the transition moment as a function of the internuclear distance $R.$ Theoretical calculations of the potentials have appeared recently \cite{jonin2002_II}.
  The choice of the potentials has to fulfill the following criteria:
\emph{i}) the upper bound state should be related to the $6p$ atomic manifold;
\emph{ii}) the selection rule for state parity $u\leftrightarrow g$ and $^{+}\leftrightarrow ^{+}$ must be obeyed \cite{herzberg};
\emph{iii}) the 
 difference between the two potential curves at the equilibrium distance of the bound state must be approximately equal to 
 $\tilde\nu_{m}.$ 
These criteria are met by the choice of the ungerade state $(3)0_{u}^{+}$ correlated with the $6p\, (^{1}\!D_{2})$ atomic limit for the  
 bound state and of the  gerade $(1)0_{g}^{+}$ state correlated with the $6s\, (^{3}\!P_{1})  $ limit as the dissociative one \cite{jonin2002_II}.
As a transition $0_{u}^{+} \rightarrow 0_{g}^{+}$ is involved, conservation of the total angular momentum enforces the additional selection rule $\Delta J =\pm 1 $ \cite{herzberg}.

As the transition moments for these states are not known, we assume that they do not vary too rapidly as a function of $ R$ 
and calculate the line shape within the 
 \emph{centroid} approximation \cite{herzberg,telli}, thus yielding
\begin{eqnarray}
 I& \propto& 
 \sum\limits_{v^{\prime}J^{\prime}} e^{-\beta E_{v^{\prime}\! J^{\prime}} }
 \left\{
 \left( J^{\prime}+1\right) \Big\vert \langle\epsilon^{\prime\prime},J^{\prime}+1 \vert
 v^{\prime},J^{\prime}\rangle\Big\vert^{2}
 \right.
 +  \nonumber\\ 
 & + &\left. 
 J^{\prime} 
 \Big\vert \langle\epsilon^{\prime\prime},J^{\prime}-1 \vert
 v^{\prime},J^{\prime}\rangle\Big\vert^{2}
 \right\} \tilde\nu^{4} 
 \label{eq:intensity2}
 \end{eqnarray}
in which the selection rule $\Delta J = J^{\prime\prime}-J^{\prime}=\pm 1$ is used. 
$\tilde \nu = \left[ \left(T_{e}^{\prime} -T_{e}^{\prime\prime}  - D_{e}^{\prime\prime} 
\right)
+ E_{v^{\prime}}
+ B_{e}^{\prime}J^{\prime} \left( J^{\prime} + 1 \right) - \epsilon^{\prime\prime}\right] $ is  the emission wave number. $E_{v^{\prime}\! J^{\prime}} = E_{v^{\prime}}
+ B_{e}^{\prime}J^{\prime} \left( J^{\prime} + 1 \right)$ is the energy of a rovibrational state. $E_{v^{\prime}}= D_{e}^{\prime} (1+\epsilon_{v^{\prime}})$ are the vibrational energy eigenvalues of the bound potential 
measured from the bottom of the potential well whereas $D_{e}^{\prime}\epsilon_{v^{\prime}}$ are the same but measured from the dissociation limit. $B_{e}^{\prime} = \hbar^{2}/2m_{r}R_{e^{\prime}}^{2}$ is the rotational constant, $ m_{r}=1.09\cdot 10^{-25}$ Kg is the reduced mass, and $R_{e^{\prime}}$ is the equilibrium internuclear distance of the bound $(3)0_{u}^{+}$ state. 
 $T_{e}^{\prime}$ and $T_{e}^{\prime\prime}$ are the values of the minimum of the potentials  
of the two states. $D_{e}^{\prime}$ and $D_{e}^{\prime\prime} $ are the depth of the potential wells. 
Though dissociative, state $(1)0_{g}^{+}$ has a weak van der Waals minimum for large 
$R$ \cite{jonin2002_II}.
$\vert v^{\prime},J^{\prime}\rangle$ is a 
rovibrational state of the bound potential. $\vert\epsilon^{\prime\prime},J^{\prime\prime}\rangle$ is a scattering state of kinetic energy $\epsilon^{\prime\prime} $ and angular momentum $J^{\prime\prime}$ in the vibrational continuum of the dissociative potential.

The exponential prefactor in Eq. \ref{eq:intensity2} accounts for the equilibrium thermal distribution of the rovibrational degrees of freedom with $\beta^{-1}= k_{\mathrm{B}}T \approx 208.5 $ cm$^{-1}.$ 

$E_{v^{\prime}}$ and $\vert v^{\prime},J^{\prime}\rangle$ are found by numerically integrating the Schr\H odinger equation for the rotationless  potential using the Numerov--Cooley finite difference scheme \cite{koonin} and 
replacing the centrifugal potential by the constant 
 $B_{e}^{\prime }J^{\prime} (J^{\prime}+1)$ \cite{herzberg}. Details will appear in a forthcoming paper. 
 For numerical purposes the theoretical potential is accurately fitted to a Morse one:
\begin{equation}
V_{b}(R) = T_{e}^{\prime} +D_{e}^{\prime} \left\{ 1 -\exp{\left[-\beta_{e^{\prime}} \left(R-R_{e^{\prime}}\right)\right]}\right\}^{2}
\label{eq:morse}\end{equation}
 with $T_{e}^{\prime}= 13860 $ cm$^{-1},$ $D_{e}^{\prime} = 1717$ cm$^{-1},$ $R_{e^{\prime }} = 3.23$ \AA, and $\beta_{e^{\prime}}R_{e^{\prime}} =6.734.$ 
The bound state accomodates up to $v^{\prime} \approx 34 $ vibrational states though only the first 10 contribute significantly to the spectrum owing to the Boltzmann factor. 

 $B_{e}^{\prime} \approx 2.47\cdot 10^{-2}$ cm$^{-1}$ yields 
a rotational temperature $\Theta_{r}\approx 3.5\cdot 10^{-2}$ K. Thus, for $T=300 $ K,  states of very high $J^{\prime}$ 
are thermally excited and their distribution is non negligible for $J^{\prime}\leq 250$ with average $\langle J^{\prime}\rangle \approx 81.$

The scattering states $\vert \epsilon^{\prime\prime},J^{\prime\prime}\rangle$ are found by numerically integrating the Schr\H odinger equation for the effective potential
\begin{equation}
V_{f\!J}(R) = V_{f}(R) + \left({\hbar ^{2}}/{2m_{r}R^{2}}\right)J^{\prime\prime}\left(J^{\prime\prime}+1\right)
 \label{eq:vfree}\end{equation}
 $V_{f}$ is the  
 potential of the $(1)0_{g}^{+}$ state 
  characterized by a shallow minimum of depth $D_{e}^{\prime\prime}\approx 217.9$ cm$^{-1}$ at $R_{e^{\prime\prime}}\approx 4.92$ \AA\ and by $T_{e}^{\prime\prime}\approx 4779.2$ cm$^{-1}$ \cite{jonin2002_II}. 
It is accurately fitted to the analytical form
$V_{f}(R) = T_{e}^{\prime\prime} + D_{e}^{\prime\prime} f(x)$ $ (x=R/R_{e^{\prime\prime}}), $
 where $f(x)$ is a HFD--B potential \cite{aziz1986}.
The Schr\H odinger equation is integrated with a Runge--Kutta 4th--order scheme with adaptive stepsize control \cite{NR}. 
The scattering 
wave functions are 
normalized to unitary incoming flux \cite{telli}. 
\begin{equation}
\psi_{\epsilon^{\prime\prime}} = \langle R\vert \epsilon^{\prime\prime},J^{\prime\prime}\rangle
 \stackrel{\scriptscriptstyle{R\rightarrow\infty}}{\longrightarrow} \left({2m_{r}}/{\pi\hbar^{2}k}\right)^{1/2}\sin{\left(k R +\eta\right)}
 \label{eq:uniflux}\end{equation}
where $\hbar^{2}k^{2}/2m_{r}=$ $\epsilon^{\prime\prime}$ and $\eta $ is the appropriate 
phaseshift.
\begin{figure}[htbp]
 \begin{center}
 \includegraphics[width=\columnwidth]{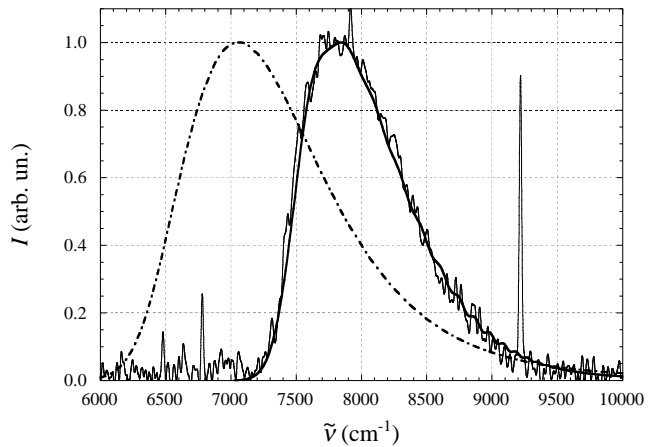}
\caption{Experimental and simulated spectra. Dash--dotted line: spectrum computed using the literature potential for the dissociative state \cite{jonin2002_II}. Solid line: spectrum calculated  using a $R^{-12}$ repulsive potential with adjustable parameters. \label{fig:ireandjon}}\end{center} 
 \end{figure}
The overlap integrals in Eq. \ref{eq:intensity2} are evaluated by spline interpolation and quadrature \cite{NR}. 
The theoretical line shape is then convoluted with the instrumental function.

In Fig. \ref{fig:ireandjon} the simulated spectrum (dash--dotted line) is compared with the experimental one.
 The shape obtained using the literature potential for the $(1)0_{g}^{+}$ state agrees only qualitatively with the experiment. It is correctly stretched towards the blue side as a consequence of the non negligible contribution of vibrational states with $v^{\prime}> 0.$ However, its position is too strongly red shifted and its width is nearly twice as large as observed. 
We conclude that the potential energy  
curves of the bound and dissociative states are too close to each other and that the dissociative potential is too steep.

The repulsive part of the lower potential can be, however, determined by inverting the line shape \cite{telli}.  We assume that the upper bound state is correctly described by the literature potential \cite{jonin2002_II} and 
that the 
lower state is described by the purely repulsive potential 
\begin{equation}
V_{rep}= A + {V_{0}}/{x^{12}} \qquad( x={R}/{R_{e^{\prime}}})
\label{eq:vrep}\end{equation} 
$R$ is scaled by the equilibrium distance of the bound state just for numerical convenience.

$A$ and $V_{0}$ are adjustable parameters to be determined by fitting the simulated spectrum to the observed one once the corresponding 
scattering wavefunctions are suitably 
computed. 
 If $A = (5315\pm 32) $ cm$^{-1}$ and $V_{0}=(760\pm 16)$ cm$^{-1},$ the simulated spectrum, (Fig. \ref{fig:ireandjon}, solid line), 
agrees perfectly with the experiment. The uncertainties on $A$ and $V_{0}$ reflect the uncertainty on the experimental determination of the spectrum position and width. 

On the blue side, tiny wiggles in the simulated spectrum reflect the contributions of vibrational states with $v^{\prime}\gg 10.$ They are not observed because of unfavorable signal--to--noise ratio and Boltzmann factor. 

 \begin{figure}[htbp]
 \begin{center}
 \includegraphics[width=\columnwidth]{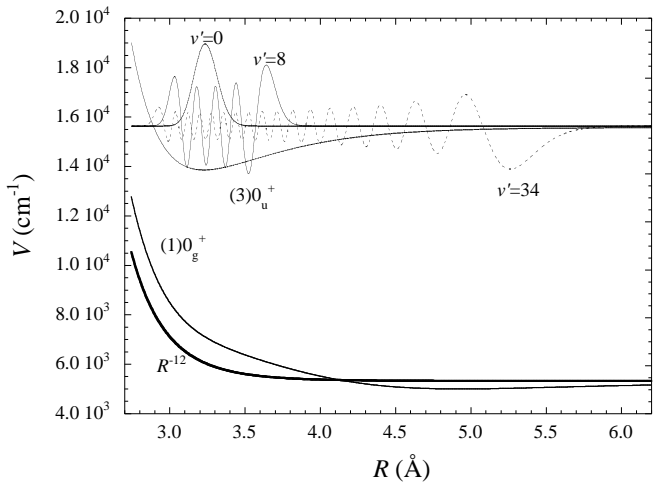}
\caption{Potentials of the $(3)0_{u}^{+}$ bound state (upper curve) and of the  dissociative $(1)0_{g}^{+}$ state (lower curves). Thin line: literature potential \cite{jonin2002_II}. Thick line: best fit potential. 
The vibrational eigenfunctions with $v^{\prime}=0,\, 8,$ and 34 are shown.
\label{fig:potentials}}\end{center} 
 \end{figure}

In Fig. \ref{fig:potentials} we compare the literature potential with that determined by the inversion procedure. The bound state potential is also shown 
with some vibrational eigenfunctions in order to visualize the coordinate range in which the Franck--Condon factors are nonnegligible. The range, in which the comparison between the potentials describing the dissociative state is reasonable, is limited to $(2.7\leq R\leq 4.5)$~\AA\ because  vibrational states with $v^{\prime}>10$  contribute little.

The difference between the best fit  potential and the theoretical one for $R=R_{e^{\prime}}$ is $\approx 880$ cm$^{-1}.$ No comparison can be made with the experimental determination of the dissociative potential of this state accomplished by REMPI techniques \cite{koe1995} because selective multiphoton excitation from the molecular ground state samples the potential for $R$ much larger than in the present case. 

We believe that the energy difference determined in this way is large enough for theoreticians to improve the calculation of the potentials and of the transition moments. However, we must stress the fact that in our analysis we have chosen to consider exact the theoretical potential of the bound state in order to modify the dissociative one.

Critical issues in our
determination of the  $(1)0_{g}^{+} $ state potential are the assumption of  validity of the centroid approximation and 
the use of the literature potential for the $(3)0_{u}^{+}$ state. We only justify these assumptions on the basis of Occam's razor. However, the assumption that emission takes place from an excimer population in thermal equilibrium 
explains the observed blue--asymmetry of the line shape,
and rules out the possibility that emission is produced by a transition from a vibrationally relaxed bound state because it  would yield a line shape of opposite asymmetry than actually 
observed.

The present analysis could be confirmed by measuring the spectrum
as a function of $T$ in order to change the distribution of the rovibrational states.
Moreover, this experiment and its consequences open up the possibility to investigate higher--lying excimer states in other rare gases that may also be of interest in astrophysics.

\begin{acknowledgments}
We acknowledge the support of D. Iannuzzi, now at Vrije Universiteit, Amsterdam, The Netherlands.
\end{acknowledgments}
\bibliography{borghesaniIRspectrum}
\end{document}